\pdfoutput=1
\documentclass[sigplan,screen,nonacm]{acmart}
\usepackage{xspace}
\usepackage{tikz}
\usetikzlibrary{arrows.meta,fit,backgrounds,positioning}
\usepackage{amssymb}
\usepackage{graphicx}
\usepackage{cleveref}
\usepackage{listings}
\usepackage{multirow}
\usepackage{enumitem}
\setlist{itemsep=1pt, topsep=3pt, parsep=0pt}
\usepackage{xcolor}
\usepackage{booktabs}

\crefname{lstlisting}{Listing}{Listings}
\Crefname{lstlisting}{Listing}{Listings}
\AtBeginDocument{%
  }

\newcommand{\tool}{\textsc{\textsc{cuVein}}\xspace}
\newcommand{\redsan}{\textsc{\textsc{RedSan}}\xspace}
\newcommand{\metric}{\textsc{\textsc{PCDepGraph}}\xspace}
\newcommand{\ncu}{\textit{NCU}}

\setcopyright{acmlicensed}
\copyrightyear{2018}
\acmYear{2018}
\acmDOI{XXXXXXX.XXXXXXX}
\acmConference[Conference acronym 'XX]{Make sure to enter the correct
  conference title from your rights confirmation email}{June 03--05,
  2018}{Woodstock, NY}
\acmISBN{978-1-4503-XXXX-X/2018/06}

\setcopyright{none}        %
\acmConference{}           %
\acmYear{}                 %
\acmDOI{}                  %
\acmISBN{}                 %
\copyrightyear{}           %
\renewcommand\keywordsname{} %
\begin{document}

\title{Reconstructing Fine-grained Memory Reuse Structure in GPU Kernels}

\author{Yanbo Zhao}
\affiliation{\institution{North Carolina State University}\city{Raleigh}\state{NC}\country{USA}}
\email{yzhao62@ncsu.edu}

\author{Yueming Hao}
\affiliation{\institution{Meta}\city{Menlo Park}\state{CA}\country{USA}}
\email{yhao@meta.com}

\author{Zecheng Li}
\affiliation{\institution{North Carolina State University}\city{Raleigh}\state{NC}\country{USA}}
\email{zli94@ncsu.edu}

\author{Mao Lin}
\affiliation{\institution{University of California, Merced}\city{Merced}\state{CA}\country{USA}}
\email{mlin59@ucmerced.edu}

\author{Shuyin Jiao}
\affiliation{\institution{North Carolina State University}\city{Raleigh}\state{NC}\country{USA}}
\email{sjiao2@ncsu.edu}

\author{Keren Zhou}
\affiliation{\institution{George Mason University}\city{Fairfax}\state{VA}\country{USA}}
\email{kzhou6@gmu.edu}

\author{Xu Liu}
\affiliation{\institution{North Carolina State University}\city{Raleigh}\state{NC}\country{USA}}
\email{xliu88@ncsu.edu}

\author{Jiajia Li}
\affiliation{\institution{North Carolina State University}\city{Raleigh}\state{NC}\country{USA}}
\email{jiajia.li@ncsu.edu}

\renewcommand{\shortauthors}{Zhao et al.}

\begin{abstract}
  Optimizing GPU data movement requires knowing which instructions access the same data and across which execution scopes reuse occurs. Existing profilers can attribute memory behavior to individual instructions but do not expose the reuse relations among them, while aliasing, dynamic indexing, and runtime parameters prevent static analysis from establishing these relations in general. Cycle-level simulation can reconstruct such relations, but at costs prohibitive for production workloads.
  We introduce two abstractions for representing this structure: \textit{Topological Distance}, which labels each dynamic memory access with the smallest execution scope (thread, warp, block, or grid) containing its most recent reuse, and \metric{}, which organizes these labels into a PC-level graph of the kernel's memory reuse structure. Defined over logical execution scopes rather than transient hardware state, the representation remains stable across runs and GPU targets while directly indicating the sharing scope that candidate optimization mechanisms must cover. Our tool, \tool{}, reconstructs \metric{} directly from optimized binaries, without sampling, at generally under 10$\times$ overhead. Across HPC and AI workloads on NVIDIA RTX 4090 and H100, optimizations guided by \metric{} eliminate up to $90.18\%$ of memory instructions and yield up to 2.84$\times$ kernel speedup; reuse-scope distributions remain consistent across platforms, supporting that the representation captures program structure rather than transient hardware behavior.
\end{abstract}


\maketitle

\section{INTRODUCTION}\label{sec:introduction}

Over the past decade, GPU compute throughput has grown far faster than off-chip memory bandwidth, making data movement a dominant bottleneck in many modern workloads~\cite{wulf1995hitting,aiandmemorywall,williams2009roofline,cao2025moe,agrawal2024taming}. Moving data efficiently depends on matching hardware mechanisms to how data is actually shared and reused during execution. This requires understanding \emph{which instructions access the same data, how those accesses are distributed across threads, and across which execution scopes reuse occurs}.

These facts, however, are what today's software stack fails to surface. GPU hardware provides mechanisms suited to reuse at different execution scopes: registers within a thread, shuffles within a warp, shared memory within a block, and the shared L2 cache across blocks. Consider a buffer placed in shared memory: it may carry communication between threads, or each thread may merely write its own elements and read them back. Both uses occupy the same memory space, but the first involves inter-thread communication while the second exhibits only thread-private reuse. This difference changes which storage mechanisms are appropriate, and telling the two apart requires relating the accessing instructions to the threads and scopes that reuse their data. Programming frameworks~\cite{ragan2013halide, triton, chen2018tvm} do not represent these reuse relations in an explicit, queryable form. Judging whether a placement matches actual use therefore requires exposing these instruction-level sharing relations.

\subsection{Fundamental Mismatch in Existing Solutions}\label{sec:introduction:mismatch}
The example generalizes to a question that recurs across memory optimizations: \emph{what is the actual reuse scope of the data an instruction touches?} Answering it requires linking each access to its reuse origin and the execution scope containing both. For optimization, this relational information must further be combined with the access's own attributes. Static analysis knows the program's thread organization, its explicit synchronization, and each access's declared memory space. When aliasing, indexing, and control flow are statically resolvable, it can prove how data is shared. In the general case, however, which accesses actually touch the same address is determined only at runtime, so the actual reuse scope cannot be established at compile time.

Dynamic observation is the natural source, and profilers~\cite{NsightCompute, rocProfiler, vtune, omniperf} are its standard form: memory-transaction volumes, cache behavior, and stall breakdowns help identify memory-bound kernels, and recent tools attribute metrics such as redundancy rates or value patterns to individual instructions~\cite{zhou2020gvprof, zhou2022valueexpert, zhao2025redsan, lin2023drgpum}. These metrics characterize individual instructions; the missing information is relational. Knowing how many accesses a shared-memory PC issues still does not say whether the data it writes is re-read by other threads or only by the thread that wrote it. Existing metrics do not link accessing PCs, reuse origins, and reuse scopes into something analyses can query together.

Cycle-level simulators~\cite{khairy2020accel} can reconstruct such relations from full execution records, but at over six orders of magnitude slowdown and typically on scaled-down inputs~\cite{avalos2021principal}. Our starting point is that the reuse facts memory-behavior analyses need, which accesses reuse whose data and within which logical execution scope, can be collected directly, without modeling hardware timing. \textbf{Organizing them, linking each access's attributes to its reuse origin and scope, produces a representation whose attributes and relations different analyses can jointly query and combine.}

\subsection{Motivation Example}\label{sec:introduction:motivation}
\Cref{code:motivation_example_l2_cache}, the key computational kernel of the Monte Carlo neutron transport proxy XSBench~\cite{Tramm:wy}, illustrates why reuse structure matters for optimization. Hardware counters report heavy global-memory traffic and a low L2 hit rate, a symptom with two structurally different explanations: the kernel may stream a large working set with little reuse, or substantial reuse may exist that the cache fails to retain. Our representation distinguishes them directly. Over 90\% of the bytes the kernel accesses re-read data whose most recent access came from a \emph{different thread block}, and the reuse volume is large relative to the accessed footprint: each interpolated field of \texttt{nuclide\_grids} touches tens of megabytes of distinct data, yet generates repeated accesses totaling hundreds of times that footprint from across the grid. Reuse at this scope can be captured only by the L2 cache shared among all SMs, yet the touched set exceeds what L2 retains under default eviction. Marking the region for preferential L2 residency through a host-side access-policy window raises the hit rate and reduces the measured global-memory traffic by 29.13\% (79.26 to 56.17\,GB), without modifying the kernel.
\begin{lstlisting}[caption={XSBench example: across different blocks, \texttt{high} and \texttt{low} derived from \texttt{nuclide\_grids} are heavily reused.}, label={code:motivation_example_l2_cache}]
__device__ void calculate_micro_xs(...){
	NuclideGridPoint * low, * high;
	//searching for target points from global nuclide_grids
  ...
    idx = grid_search_nuclid(...)
    @low@ = @&nuclide_grids[... idx ...]@;
  ...
	@high@ = @low@ + 1;
	...
	xs_vector[0] = @high->total_xs@ - f * (@high->total_xs@ - @low->total_xs@);
	... //four more fields interpolated likewise
}
\end{lstlisting}

\subsection{Our Approach}\label{sec:introduction:approach}
We characterize memory reuse through the GPU's logical execution hierarchy. Threads, warps, blocks, and grids form nested scopes defined by the program and its launch configuration, with each scope corresponding to hardware mechanisms suited to serving reuse within it. We classify each dynamic reuse by the smallest execution scope containing both accesses. This captures how data is shared across execution contexts while abstracting away transient physical residency. Aggregating these scope-labeled reuse relations by PC yields a compact representation that connects sharing behavior to the responsible instructions and potential optimization mechanisms, with a size independent of dynamic trace length.

We realize this design as two constructs. \textit{Topological Distance} labels each dynamic memory access with the smallest execution scope (thread, warp, block, or grid) containing its most recent reuse. The Program Counter Dependency Graph (\metric{}) organizes these observations into a graph whose nodes are memory-access PCs annotated with access attributes (type, memory space, and per-execution efficiency distributions), and whose edges link each access to its reuse origin, labeled with Topological Distance and weighted by traffic. Because both are defined over logical execution scopes rather than transient hardware state, the representation remains directly interpretable by programmers and compilers.

\metric{} supports multiple analyses by composing node attributes with reuse relations. Node attributes alone support per-instruction access-efficiency analysis (coalescing, broadcast, bank behavior). Joining node attributes with scope-labeled edges supports reuse-pair analysis: which accesses are redundant and which execution scope contains their reuse. Groups of nodes and their relations expose sharing structure: closed subgraphs identify instruction groups whose reuse relations remain within the group, which in our softmax case study separates two logical buffers within a single allocation (\Cref{sec:case_studies:ai_redundancy}). A taxonomy of memory-behavior patterns is then expressible as mechanical rules over these attributes and relations, relating observed structure to the hardware mechanisms suited to serve it and narrowing the candidate optimizations (\Cref{sec:recognized_patterns}).

Our tool, \tool{}, makes the representation obtainable in practice: lightweight binary instrumentation and a parallel offline analyzer reconstruct \metric{}, without sampling, directly from fully optimized binaries, without source modification or specialized hardware support, with end-to-end overhead generally under 10$\times$.

\subsubsection*{Contribution Summary}
\begin{itemize}[leftmargin=*]
    \item We introduce \textit{Topological Distance} and \metric{}, a runtime representation of GPU memory reuse structure defined over execution scopes. Its node attributes and scope-labeled reuse relations jointly express the dynamic facts that static analysis must conservatively forgo and aggregate counters cannot express. (\Cref{sec:topological_distance,sec:pcdg})
    \item We define a taxonomy of memory-behavior patterns as mechanical rules over the representation, spanning the instruction, reuse-pair, and data-object levels and relating observed structure to candidate hardware mechanisms. (\Cref{sec:recognized_patterns})
    \item We implement \tool{}, which reconstructs \metric{} from optimized binaries, without sampling, at generally under 10$\times$ overhead. We validate the representation for fidelity (against hardware counters and analytical microbenchmarks), stability (across runs and GPU platforms), and actionability: guided fixes eliminate up to 90.18\% of memory instructions and reach up to 2.84$\times$ kernel speedup. (\Cref{sec:overall_workflow,sec:evaluation})
\end{itemize}

\section{RELATED WORK}\label{sec:related_work}
Prior work on GPU performance diagnosis has mainly advanced along two complementary directions. One focuses on memory behavior analysis, characterizing inefficient accesses, redundancy, and locality loss from the perspective of memory transactions and data objects. The other emphasizes control/data/value flow analysis, explaining how dependencies, value propagation, and SIMT semantics shape performance behavior.

\noindent\textbf{Memory Behavior Analysis:}
Memory behavior recognition is a key approach for observing and diagnosing practical GPU inefficiencies, and instruction-level metrics are usually more actionable than kernel-level aggregates because developers can directly map abnormal transactions back to concrete instructions and data structures.
  \textit{DrGPUM}~\cite{lin2023drgpum} observes data-object level memory inefficiency patterns through a runtime map of objects.
  \textit{MDM}~\cite{wang2020mdm} measures memory divergence to detect poor spatial locality under SIMT execution.
  \textit{DrGPU}~\cite{hao2023drgpu} adopts a top-down methodology that attributes GPU inefficiencies to hierarchical bottleneck categories.
  \textit{Accel-Sim}~\cite{khairy2020accel} enables detailed modeling of GPU cache and memory behavior through simulation.
Our work, by contrast, is memory-centric: it describes data flow across threads directly from observed memory reuse, without requiring hardware simulation.

\redsan{}~\cite{zhao2025redsan} pinpoints only redundant memory transactions. This is a narrow subset of the patterns our work detects (Sec.~\ref{sec:recognized_patterns}); more fundamentally, a redundancy verdict discards \emph{which} instruction the reuse repeats and \emph{which} execution-scope boundary it crosses---precisely the information that determines the appropriate fix. Our design is also faster: by replacing \redsan{}'s serial hashmap shadow memory with a direct-mapped, lock-free layout (Sec.~\ref{sec:overall_workflow}), we reduce the tens-fold slowdown typical of shadow-memory GPU analysis to generally under
$10\times$.

\noindent\textbf{Locality Metrics and Reuse Distance:}
Reuse distance (LRU stack distance)~\cite{ding2003predicting} is the canonical machine-\hspace{0pt}independent locality metric on CPUs: the number of distinct elements between two accesses to the same datum directly predicts LRU cache behavior, and \textit{CUDAAdvisor}~\cite{shen2018cudaadvisor} brings its measurement to GPUs via LLVM instrumentation. The metric presumes a single sequential access stream, in which the intervening count reflects the program's own locality. On a GPU, tens of thousands of concurrent threads interleave into that stream, so the count is dominated by whatever unrelated accesses the parallelism happens to interpose, and, since the interleaving is decided by warp scheduling, it varies from run to run for the same kernel. Collection cost compounds the problem: a search-structure update on every access already pushes CPU tools toward sampling. These observations motivate Topological Distance, which characterizes reuse by execution scope rather than temporal distance (\Cref{sec:topological_distance}).

\noindent\textbf{Control/Data/Value Flow Analysis:}
GPU SIMT semantics and hierarchical memory spaces pose additional challenges to traditional data flow analysis.
  \ncu{}~\cite{NsightCompute} provides a static register dependency tree reflecting the compiler's instruction selection and scheduling.
  \textit{GPA}~\cite{zhou2021gpa} uses data flow analysis over GPU PC samples to attribute instruction stalls to root causes.
  \textit{HPCToolkit}~\cite{adhianto2010hpctoolkit} has been extended to support GPU profiling through PC sampling and binary analysis of CUDA binaries, enabling metric attribution across heterogeneous CPU--GPU contexts.
  \textit{GVProf}~\cite{zhou2020gvprof} and \textit{ValueExpert}~\cite{zhou2022valueexpert} perform value-flow analysis that tracks how value patterns propagate to memory operands to identify value-related inefficiencies; foundational CPU-side predecessors include \textit{DeadSpy}~\cite{chabbi2012deadspy} and \textit{RedSpy}~\cite{wen2017redspy}.
  Compiler transformations that exploit uniform or redundant SIMT structure~\cite{wang2017decoupled, yeh2020dimensionality, saumya2022darm} are natural downstream consumers of the reuse structure \metric{} exposes.

\section{METHODOLOGY}\label{sec:methodology}

In this section, we first define \textit{Topological Distance} as an instruction-level abstraction of fine-grained memory reuse across GPU execution scopes. We then introduce \metric{}, which aggregates these dynamic reuse observations at the PC level into a structured representation that preserves both reuse origin and reuse scope. Building on \metric{}, we further define analysis attributes and representative memory-behavior patterns, showing how different access behaviors manifest as different reuse structures and efficiency implications on GPU hardware.

\subsection{Topological Distance}\label{sec:topological_distance}

GPU kernels execute over a hierarchical organization of threads, warps (32 threads sharing a SIMT schedule), blocks, and the kernel-wide grid, whose nested levels form natural scopes over which data reuse can arise; we adopt NVIDIA terminology throughout, other vendors exposing analogous scopes (e.g., AMD's wavefronts and workgroups).
Topological Distance is an abstraction of the fine-grained reuse cost of an individual memory instruction. For each dynamic memory instruction instance, we locate the most recent prior access to the same tracked memory address, and assign the current instruction a Topological Distance according to the smallest execution scope that contains both accesses. Intuitively, Topological Distance reflects how far data must travel through the GPU execution hierarchy to serve the current instruction from prior reuse.
We define Topological Distance over logical execution scopes rather than dynamic physical residency, because logical scopes are stable, interpretable, and more actionable for optimization (discussed below).

We define Topological Distance using the following six levels, which later serve as edge attributes in our \metric{} (\Cref{sec:pcdg}); \Cref{code:topo_distance_example} illustrates each level with a minimal annotated access sequence:

\begin{itemize}[leftmargin=*]
\item \textbf{Cold Miss}: the current instruction accesses a memory region for which no prior access exists in the current analysis context.
\item \textbf{Intra-Thread}: the most recent prior access to the tracked memory region comes from the same thread.
\item \textbf{Intra-Warp-Same-Inst}: the most recent prior access comes from another thread in the same warp during the same dynamic instruction issue. This case corresponds to warp-uniform accesses that can be naturally served by broadcast-style hardware behavior.
\item \textbf{Intra-Warp-Diff-Inst}: the most recent prior access comes from another thread in the same warp, but from a different dynamic instruction issue.
\item \textbf{Intra-Block}: the most recent prior access comes from a different warp in the same block.
\item \textbf{Intra-Grid}: the most recent prior access comes from a different block in the same grid.
\end{itemize}
\begin{figure}[t]
\begin{lstlisting}[caption={A minimal kernel fragment illustrating every Topological Distance level on a global array \texttt{a} (\texttt{tid}: thread index within a block). Each access is annotated with the label of (most of) its lanes, assuming a power-of-two block size of at least 64.}, label={code:topo_distance_example}]
int gid = blockIdx.x * blockDim.x + tid;
float s = a[gid];        // Cold Miss
s += a[gid];             // Intra-Thread
s += a[gid & ~31];       // Intra-Warp-Same-Inst
s += a[gid ^ 1];         // Intra-Warp-Diff-Inst
s += a[gid ^ 32];        // Intra-Block
s += a[gid ^ blockDim.x];// Intra-Grid
\end{lstlisting}
\vspace{-0.4cm}
\end{figure}

These levels follow the GPU execution hierarchy rather than any particular hardware generation, so scopes introduced by hardware evolution extend the taxonomy naturally (e.g., thread-block clusters add a cluster level between \textbf{Intra-Block} and \textbf{Intra-Grid}); our current implementation targets the four classical scopes.

Defining \textit{Topological Distance} over logical scopes rather than physical residency is deliberate, for three reasons. \textit{First, it remains effective}: a larger \textit{Topological Distance} generally corresponds to a longer physical data-movement path, from thread-private registers, to SM-local storage, to inter-SM L2, and ultimately to off-chip global memory. \textit{Second, it improves interpretability and actionability}, because thread, warp, block, and grid are stable execution scopes that programmers and compilers can reason about directly: a thread's warp and block membership is fixed at launch, and each block is bound to a single Streaming Multiprocessor (SM) for its lifetime.\footnote{Rare block rebinding (e.g., under debugging or UVM page faults) does not materially affect scope attribution.} \textit{Third, it is more reliable}: physical residency of data and thread are difficult to characterize accurately, because cache-sharing effects depend on hardware management and are neither deterministic nor guaranteed by the programming model.

\begin{figure}[htbp]
    \centering
    \vspace{-0.2cm}
    \resizebox{\linewidth}{!}{\begin{tikzpicture}[
  x=1cm, y=1cm,
  pc/.style={draw=black!70, rounded corners=2pt, fill=black!78, text=white,
             font=\fontsize{5.5}{6.2}\selectfont\ttfamily, align=center,
             inner xsep=3pt, inner ysep=2.5pt},
  coldnode/.style={circle, draw=blue!55!black, fill=blue!65, text=white,
             font=\fontsize{5.5}{6.2}\selectfont, align=center, inner sep=1.5pt},
  grp/.style={draw, dashed, rounded corners=4pt, inner sep=5pt},
  glab/.style={font=\fontsize{5.5}{6}\selectfont\itshape, text=black!60},
  et/.style={-{Stealth[length=4.5pt]}, red!75!black},
  ew/.style={-{Stealth[length=4.5pt]}, yellow!45!orange},
  eb/.style={-{Stealth[length=4.5pt]}, orange},
  eg/.style={-{Stealth[length=4.5pt]}, green!55!black},
  es/.style={-{Stealth[length=4.5pt]}, magenta!70, densely dashed},
  ec/.style={-{Stealth[length=4.5pt]}, blue!60, densely dashed},
  w1/.style={line width=0.5pt}, w2/.style={line width=1.0pt}, w3/.style={line width=1.7pt},
]
\node[pc] (n790) at (0,0)     {0x790\\st.shared 4B};
\node[pc] (n890) at (0,-1.15) {0x890\\ld.shared 4B};
\node[pc] (n980) at (0,-2.30) {0x980\\st.shared 4B};
\node[pc] (na80) at (0,-3.45) {0xa80\\ld.shared 4B};
\node[pc] (n860) at (2.05,0)     {0x860\\st.shared 4B};
\node[pc] (n880) at (2.05,-1.15) {0x880\\ld.shared 4B};
\node[pc] (na30) at (2.05,-2.30) {0xa30\\st.shared 4B};
\node[pc] (na70) at (2.05,-3.45) {0xa70\\ld.shared 4B};
\node[pc] (n720) at (4.15,-0.9)  {0x720\\ld.global 4B};
\node[pc] (na50) at (4.15,-2.05) {0xa50\\ld.global 4B};
\node[pc] (n6b0) at (6.05,-0.9)  {0x6b0\\ld.global 2B};
\node[pc] (n6f0) at (6.05,-2.05) {0x6f0\\ld.global 4B};
\node[pc] (nba0) at (6.05,-3.20) {0xba0\\st.global 4B};
\coordinate (g3padR) at (5.15,-1.5);
\coordinate (g3padL) at (3.55,-1.5);
\node[coldnode] (cold) at (4.9,0.85) {untracked\\memory};

\begin{scope}[on background layer]
\node[grp, draw=green!55!black, fill=green!8,  fit=(n790)(na80)] (g1) {};
\node[grp, draw=blue!55,        fill=blue!6,   fit=(n860)(na70)] (g2) {};
\node[grp, draw=violet!70,      fill=violet!7, fit=(n720)(na50)(g3padR)(g3padL)] (g3) {};
\end{scope}
\node[glab, anchor=north] at (g1.south) {Group 1};
\node[glab, anchor=north] at (g2.south) {Group 2};
\node[glab, anchor=north] at (g3.south) {Group 3};

\draw[et,w2] (na80.north) -- (n980.south);
\draw[et,w2] (n980.north) -- (n890.south);
\draw[et,w2] (n890.north) -- (n790.south);
\draw[et,w1] (na70.east) to[bend right=40] (na30.east);
\draw[et,w1] (na30.east) to[bend right=40] (n880.east);
\draw[et,w1] (n880.east) to[bend right=40] (n860.east);
\draw[ew,w2] (na70.north) -- (na30.south);
\draw[ew,w2] (n880.north) -- (n860.south);
\draw[eb,w2] (na70.165) to[bend left=28] (n880.200);
\draw[eb,w2] (na70.150) to[bend left=45] (n860.205);
\draw[eb,w3] (na30.165) to[bend left=28] (n880.160);
\draw[eb,w2] (na30.150) to[bend left=45] (n860.190);
\draw[eb,w2] (n880.150) to[bend left=30] (n860.172);
\draw[es,w3] (n720.15) .. controls +(0.55,0.22) and +(0.55,-0.22) .. (n720.-15);
\draw[es,w3] (na50.15) .. controls +(0.55,0.22) and +(0.55,-0.22) .. (na50.-15);
\draw[eg,w2] (na50.west) to[bend left=35] (n720.west);
\draw[eg,w2] (n720.south) to[bend left=30] (na50.north);
\draw[eg,w2] (n720.165) .. controls +(-0.45,0.5) and +(0.45,0.5) .. (n720.60);
\draw[ec,w3] (n790.north) to[bend left=12] (cold.west);
\draw[ec,w3] (n860.north) to[bend left=8]  (cold.west);
\draw[ec,w1] (n720.north) -- (cold.south west);
\draw[ec,w3] (n6b0.north) to[bend right=8] (cold.south east);
\draw[ec,w3] (n6f0.east) to[bend right=45] (cold.east);
\draw[ec,w3] (nba0.east) to[bend right=55] (cold.east);
\begin{scope}[every node/.style={font=\fontsize{5.5}{6}\selectfont, anchor=west, inner sep=1pt}]
\draw[et,w2] (-0.65,-4.45) -- (-0.33,-4.45);  \node at (-0.28,-4.45) {intra-thread};
\draw[es,w2] (1.8,-4.45) -- (2.12,-4.45);     \node at (2.17,-4.45) {intra-warp-same-inst};
\draw[eg,w2] (4.85,-4.45) -- (5.17,-4.45);    \node at (5.22,-4.45) {intra-grid};
\draw[eb,w2] (-0.65,-4.77) -- (-0.33,-4.77);\node at (-0.28,-4.77) {intra-block};
\draw[ew,w2] (1.8,-4.77) -- (2.12,-4.77);   \node at (2.17,-4.77) {intra-warp-diff-inst};
\draw[ec,w2] (4.85,-4.77) -- (5.17,-4.77);  \node at (5.22,-4.77) {cold miss};
\end{scope}
\end{tikzpicture}}
    \vspace{-0.4cm}
    \caption{An example \metric{} from the \textit{soft\_max\_f32} kernel of \textit{llama.cpp/softmax}. The blue circle node represents first accesses to untracked memory regions (\textit{Cold Miss}); edge colors encode the Topological Distance of each reuse (see legend), and edge width is logarithmically scaled by edge weight to maintain readability over a wide dynamic range.
    } 
    \vspace{-0.4cm}
    \Description{A program-counter dependency graph where nodes are PCs and edges indicate reuse relationships, including cold misses.}
    \label{fig:dependency_graph_example}
\end{figure}
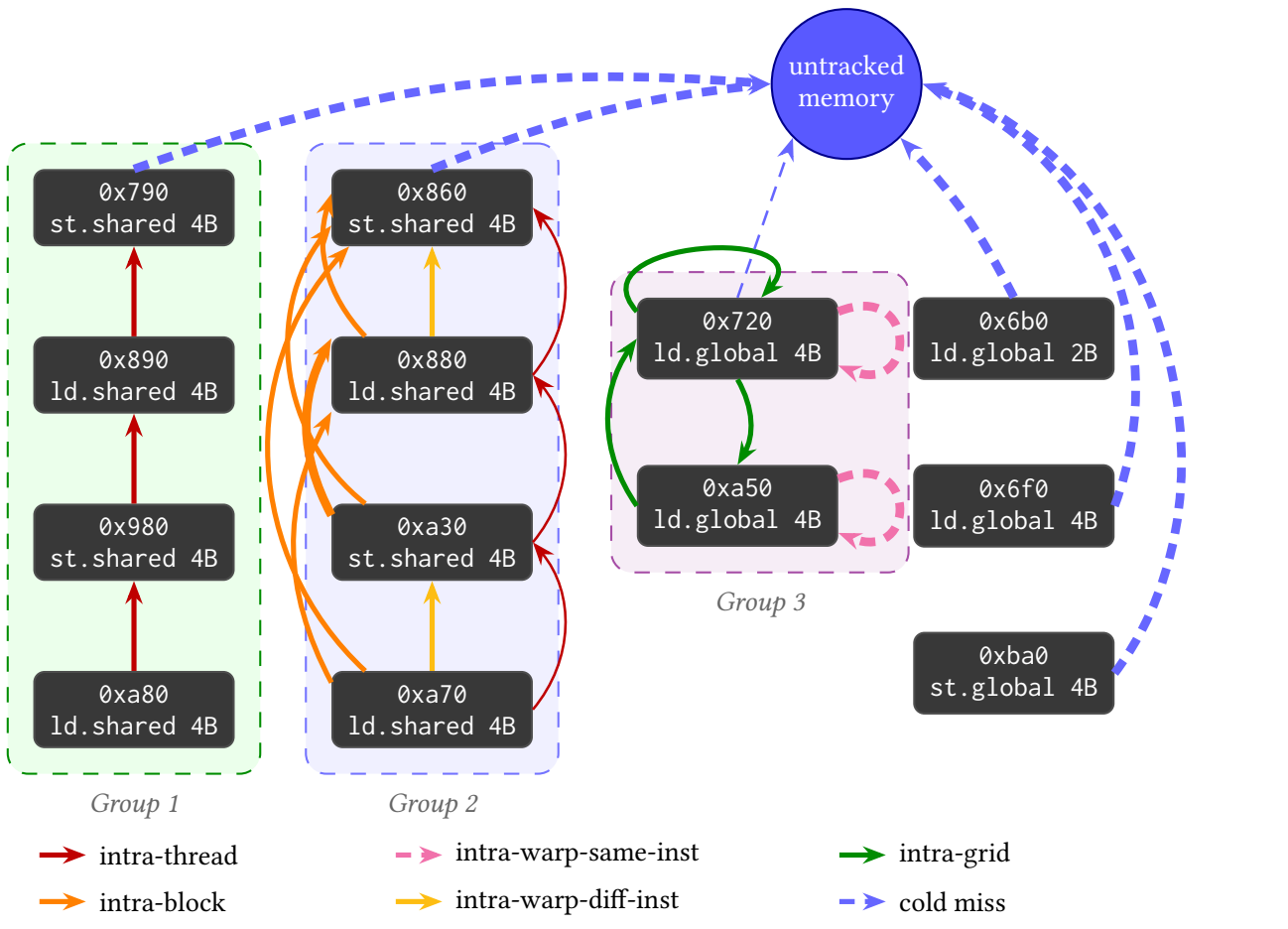

\subsection{\metric{}}\label{sec:pcdg}

\subsubsection{Definition}

We aggregate instruction-level Topological Distance observations at the PC level into \metric{}, a graph structure. Intuitively, each node in \metric{} represents a memory-access PC, and each edge summarizes where the data used by that PC most recently came from. Formally, \metric{} is a directed multi-attribute graph $G = (V, E, \Phi, w)$, where:

\begin{itemize}[leftmargin=*]
    \item \textbf{$V$}: a set of nodes, where each $v \in V$ represents a Program Counter (PC) corresponding to a memory instruction. Each node carries instruction attributes including access type (\textit{Read}, \textit{Write}, or \textit{Atomic}), memory space (\textit{Shared}, \textit{Global}, or \textit{Local}), and access \textit{size}. We introduce a special node, ``untracked memory addresses'', representing first accesses to memory regions and encoding the \textit{Cold Miss} level of Topological Distance.\footnote{Asynchronous copies (\texttt{LDGSTS}), which touch two memory spaces, are handled in \Cref{sec:overall_workflow}.}
    
    \item \textbf{$E$}: a multiset of directed edges $e = (PC_{\mathit{cur}} \rightarrow PC_{\mathit{prev}})$. An edge exists when the current access issued at $PC_{\mathit{cur}}$ touches a tracked memory region whose most recent prior access was issued at $PC_{\mathit{prev}}$; note that each edge points from an access \emph{back} to its reuse origin, the \emph{reverse} of the producer-to-consumer orientation of dataflow graphs.
    
    \item \textbf{$\Phi$}: a labeling function $\Phi: E \rightarrow L$, where $L = \{\textit{Thread},$ $\textit{Warp}_{\mathit{same\_inst}},$ $\textit{Warp}_{\mathit{diff\_inst}},$ $\textit{Block},$ $\textit{Grid},$ $\textit{ColdMiss}\}$. For each edge, $\Phi(e)$ records one of the six levels of Topological Distance associated with the current access.
    
    \item \textbf{$w$}: a weight function $w: E \rightarrow \mathbb{R}_{>0}$, where $w(e)$ denotes the data volume accumulated on edge $e$.
\end{itemize}

\Cref{fig:dependency_graph_example} shows an example of \metric{}. The graph contains 13 memory-access PCs plus the special \textit{Cold Miss} node, connected by 24 edges, 6 of them cold-miss edges. Edge colors indicate different Topological Distance labels, and edge width represents the corresponding edge weight. Self-loops are allowed, because repeated executions of the same PC may access the same tracked memory region. 

For a subgraph $G'=(V',E') \subseteq G$, we call $G'$ a \textit{closed subgraph} if, for every node $v \in V'$, all non-\textit{Cold Miss} edges of $v$ remain within $V'$; \Cref{fig:dependency_graph_example} contains three, and reuse happens only within them.

Because reuse edges follow addresses, the nodes of a closed subgraph are exactly the instructions that share a set of memory regions, in practice a data object or a tile of one. Its cold-miss edge weight is thus the object's unique footprint, and its total edge weight the object's actual access traffic, decomposed by reuse scope. This attribution operates on \emph{logical} objects rather than allocations: the two shared-memory subgraphs in \Cref{fig:dependency_graph_example} are two logical buffers residing in the same physical allocation (\Cref{sec:case_studies:ai_redundancy}). Hardware counters offer no such object-level attribution, and unlike allocation-tracking profilers~\cite{lin2023drgpum}, it requires neither allocation interception nor source information.

\metric{} turns dynamic, instruction-level reuse observations into a compact graph representation that preserves both reuse origin and reuse scope, revealing directly from observed reuse (without hardware simulation) which instructions carry reuse-intensive traffic, which are dominated by first-touch accesses, and which repeatedly obtain data from more distant execution scopes. As a result, the graph is determined by the program's memory access behavior and grid configuration, independent of the executing hardware up to a single schedule-dependent tie-break at scope boundaries, which we quantify in \Cref{sec:evaluation:fidelity}.

\metric{} is intentionally designed for GPU programs rather than as a generic cross-platform memory-behavior graph: its interpretability rests on the stability of GPU execution scopes established in \Cref{sec:topological_distance}. By contrast, on CPUs, unless threads are explicitly pinned, the OS scheduler may migrate threads across physical cores, weakening fixed hardware-scope interpretations of access locality.

\begin{figure}[h]
    \centering
    \vspace{-0.2cm}
    \includegraphics[width=0.62\linewidth]{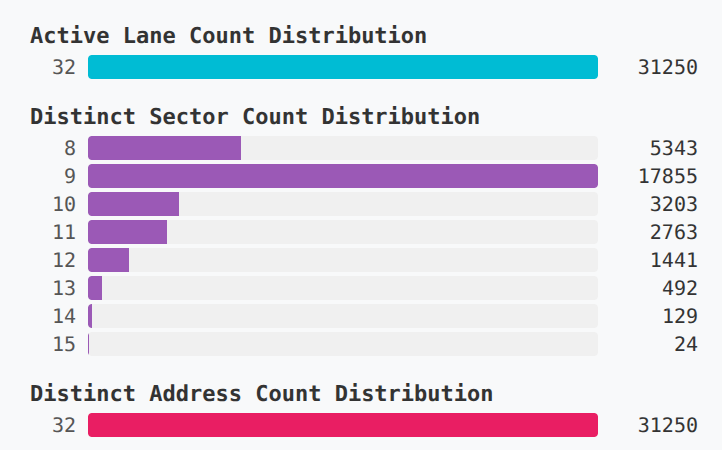}
    \vspace{-0.2cm}
    \caption{Example distributions of one 8-byte-access PC: a perfectly coalesced execution would touch 8 sectors; the collected distribution shows a scattered access pattern.}
    \label{fig:distributions}
    \vspace{-0.4cm}
\end{figure}

\subsubsection{Analysis Attributes} 

\metric{} captures reuse structure, but that alone is not sufficient for diagnosing memory efficiency: two PCs may exhibit similar reuse topology while differing substantially in coalescing quality or active-lane participation. We therefore augment each PC node with dynamic attributes characterizing how efficiently the instruction executes, kept as full distributions rather than the averages hardware counters report.

\noindent\textbf{Dynamic access distributions.}
For each node $v \in V$, we maintain three dynamic distributions (\Cref{fig:distributions}) collected over all executions of the corresponding PC:

\begin{itemize}[leftmargin=*]
    \item \textbf{Active-lane count distribution} ($Lane_v$): the number of active lanes in each warp-level execution, reflecting control-flow divergence and boundary effects.

    \item \textbf{Distinct-address count distribution} ($Addr_v$): the number of unique element addresses accessed per execution, capturing intra-sector duplication such as broadcast-like accesses or partial-lane contention.

    \item \textbf{Distinct-sector count distribution} ($Sector_v$): the number of distinct memory sectors touched per execution, directly reflecting coalescing quality.
\end{itemize}

For shared-memory PCs, these distributions are reinterpreted at bank-conflict granularity: shared memory has no sectors, so $Addr_v$ becomes a distinct 32-bit-word count, and in place of $Sector_v$ we record a per-instance \textbf{wavefront count distribution} ($Wave_v$). For each warp-level execution, every active lane's access footprint is mapped to its 32-bit words, duplicates are removed, each word is assigned to one of the 32 banks ($(\mathit{addr} \gg 2) \,\&\, 31$ under the 32-bank, 4-byte-wide organization of NVIDIA shared memory), and the wavefront count is the maximum number of distinct words that fall on any single bank, so broadcasting one word to any number of lanes adds no cost; wide accesses that move more than 4 bytes per lane are processed by the hardware in fixed lane groups (e.g., a 128-bit store handles eight lanes per group), and their count sums the per-group maxima. This distribution underlies the \textit{Bank Conflict} pattern (\Cref{sec:recognized_patterns}).

Together, these distributions distinguish whether low memory efficiency arises from scattered addressing, intra-sector duplication, or limited lane utilization. The lane, sector, and address attributes are collected on-the-fly by warp-level primitives in the \textit{Trace Collector}, while the word-granularity wavefront counts are derived by the \textit{Offline Analyzer} from the recorded per-lane addresses (\Cref{sec:overall_workflow}).

\noindent\textbf{Weighted Out-Degree.}
For a node $v \in V$, let $E_{out}(v)$ and $E_{in}(v)$ denote the sets of outgoing and incoming edges of $v$, respectively. For a Topological Distance label $l \in \mathcal{L}$, we define the \textit{Weighted Out-Degree} of $v$ at scope $l$ and the \textit{Total Weighted Out-Degree} of $v$ as
\begin{gather*}
\mathcal{W}_{out}(v, l) = \textstyle\sum_{e \in E_{out}(v),\, \Phi(e)=l} w(e), \\
\mathcal{W}_{total}(v) = \textstyle\sum_{l \in \mathcal{L}} \mathcal{W}_{out}(v, l).
\end{gather*}
Intuitively, $\mathcal{W}_{out}(v,l)$ measures how much data volume issued by PC $v$ obtains its most recent reuse from scope $l$, while $\mathcal{W}_{total}(v)$ measures the total outgoing data volume associated with $v$ across all Topological Distance labels.
We focus on outgoing edges because, under the current-to-previous edge direction in \metric{}, they directly characterize the reuse sources of accesses issued by PC $v$.

\noindent\textbf{Reuse Intensity.}
Beyond per-node access-efficiency attributes, we define a subgraph-level metric, \textit{Reuse Intensity} ($RI$), to summarize how strongly a set of PCs depends on reused data rather than first-touch accesses. For a closed subgraph
$G' \subseteq G$ with edge multiset $E(G')$, we define
\[
RI(G') = \frac{\sum_{e \in E(G')} w(e)}{\sum_{e \in E(G'),\,\Phi(e)=\textit{ColdMiss}} w(e)}.
\]
Intuitively, $RI$ increases when a subgraph has substantial reuse traffic relative to cold-miss footprint. Under the data-centric interpretation of closed subgraphs above, $RI(G')$ is the traffic amplification of the corresponding data object relative to its unique footprint, making it particularly useful for identifying hot memory regions.

\subsection{Identifiable Memory Behavior Patterns}\label{sec:recognized_patterns}

In this section, we present representative memory-behavior patterns identifiable in \metric{} and organize them by their most relevant execution scope: \textbf{Thread}, \textbf{Warp}, \textbf{Block\slash Grid}, and \textbf{Others}. These patterns cover both common memory behaviors and potentially inefficient ones. Each scope has hardware mechanisms suited to serving its reuse (registers for intra-thread, warp shuffles for intra-warp, shared memory or the SM-local L1 for intra-block, and the SM-shared L2 for intra-grid); reuse served by a mechanism that does not match its scope leaves performance on the table, and it is precisely such mismatches that the inefficient patterns below diagnose. Together, they provide a practical vocabulary for interpreting \metric{} and connecting observed reuse structure to likely optimization implications.

\noindent\textbf{I. Block/Grid level.} \mbox{}
\label{sec:block-grid-level}

\noindent \underline{\textit{Hierarchical Reuse.}} \label{pattern:hierarchical_reuse}
For a subgraph $G' \subseteq G$, the reuse ratio at scope $l$ is
\[
r(l)=\frac{\sum_{v \in V(G')} \mathcal{W}_{out}(v, l)}{\sum_{v \in V(G')} \widehat{\mathcal{W}}(v)},
\]
where $\widehat{\mathcal{W}}(v)$ is $\mathcal{W}_{total}(v)$ except its $Warp_{same\_inst}$ and \textit{ColdMiss} components, to exclude non-reused and broadcast volumes (\Cref{sec:topological_distance}).
We say that $G'$ exhibits a dominant reuse pattern at scope $l$ if $r(l)$ exceeds a threshold $\delta$ (e.g., $\delta=0.8$ in our experiments). When $l=Grid$, the subgraph is dominated by intra-grid reuse, indicating reuse across block boundaries and potential pressure on the L2 cache. \label{pattern:intra-grid} When $l \in \{Thread, Warp_{diff\_inst}, Block\}$, the subgraph is dominated by reuse within a block, suggesting potential opportunities for shared-memory promotion. \label{pattern:intra-block}

\noindent\textbf{II. Warp level.} \mbox{} 

\noindent \underline{\textit{Poor Coalescing.}} \label{pattern:low_coalescing}
For a node $v$, let $E(Sector_v)$ denote the weighted average distinct sector count derived from the distribution $Sector_v$ defined in \Cref{sec:pcdg}. The theoretical optimal sector count is
\[
S_{opt}(v)=\left\lceil \frac{Lane_v \cdot \mathit{AccSize}}{\mathit{SecSize}} \right\rceil,
\]
where $Lane_v$ is the active-lane count, $\mathit{AccSize}$ is the access-size attribute of the node in \metric{}, and $\mathit{SecSize}$ is the hardware sector size (32 bytes on NVIDIA GPUs). 
We quantify coalescing quality as
\[
R_{coal}(v)=\frac{E(Sector_v)}{S_{opt}(v)};
\]
$R_{coal}(v)>1$ means the active lanes touch more sectors than theoretically necessary (sparse or unaligned accesses that leave fetched sector data unused); in the worst case, each lane touching a different sector, the ratio approaches $\mathit{SecSize}/\mathit{AccSize}$ (8 for 4-byte accesses).

\noindent \underline{\textit{Broadcasting/Multicasting/Atomic Contention.}} \label{pattern:broadcasting}
A node $v$ is classified as \textit{Broadcasting} when it satisfies: (1) $Type(v)=\textit{Read}$, (2) $Lane_v > 1$, and (3) $\text{Mode}(Addr_v)=1$, i.e., multiple active lanes read the same address in the same instruction instance, a warp-level broadcast that hardware mechanisms such as the constant-cache path serve efficiently. If instead $\text{Mode}(Addr_v) \in (1,Lane_v)$, $v$ is classified as \textit{Multicasting}, or as \textit{Atomic Contention} if $Type(v)=\textit{Atomic}$.

\noindent \underline{\textit{Bank Conflict.}} \label{pattern:bank_conflict}
The hardware serializes a shared-memory instruction instance with wavefront count $k$ (the $Wave_v$ attribute of \Cref{sec:pcdg}) into $k$ shared-memory cycles. We define the conflict level of a shared-memory node $v$ as
\[
R_{conf}(v) = \frac{E(Wave_v)}{E(Wave^{ideal}_v)},
\]
where the ideal count of an instance, its bandwidth floor, is $\lceil \text{footprint words}/32 \rceil$ accumulated over the same lane groups. A conflict-free access pattern yields that $R_{conf}(v)=1$; the excess over $1$ measures pure serialization waste, approaching the warp width in the worst case (all active lanes addressing distinct words in a single bank). The computed wavefronts are validated against the hardware's shared-memory wavefront counters in \Cref{sec:evaluation:fidelity}.

\noindent\textbf{III. Thread level.} \mbox{} 

\noindent \underline{\textit{Redundant Access.}}\label{pattern:redundant} 
A \textit{Redundant Access} is defined on two different nodes. For an edge $e=(u \to v)$ in \metric{}, $u$ denotes the current PC and $v$ denotes the most recent prior PC. We identify the following three cases:

\noindent$\bullet$\textbf{RAR:} $\forall e \in E_{out}(u)$, if $\Phi(e)=Thread$, $Type(u)=Read$, and $Type(v)=Read$, then $u$ is a redundant read: it re-reads values already read by the same thread.

\noindent$\bullet$\textbf{RAW:} $\forall e \in E_{out}(u)$, if $\Phi(e)=Thread$, $Type(u)=Read$, and $Type(v)=Write$, then $u$ is a redundant read: it re-reads values already produced by the same thread.

\noindent$\bullet$\textbf{WAW:} $\forall e \in E_{in}(v)$, if $\Phi(e)=Thread$, $Type(u)=Write$, and $Type(v)=Write$, then $v$ is a redundant write: every value it produces is overwritten by the same thread with no reuse in between, making $v$ a dead store.

A redundant PC indicates a strong candidate for REG promotion, but the REG budget should be considered to avoid decreasing occupancy.

\noindent \underline{\textit{Abuse SMEM.}}\label{pattern:thread_private_shared}
A closed subgraph $G'$ is identified as exhibiting only thread-private SMEM use when all nodes in $G'$ are shared-memory instructions and all edges in $G'$ are intra-thread reuse edges.
This pattern indicates that the data placed in shared memory is reused only within individual threads, without inter-thread sharing in the subgraph.

\noindent\textbf{IV. Others.} \mbox{} 

\noindent \underline{\textit{Streaming Access.}}\label{pattern:streaming} 
A node $v$ is identified as exhibiting the \textit{Streaming Access} pattern when its outgoing data flow in \metric{} is dominated by \textit{ColdMiss} edges, i.e., when the ratio $\mathcal{W}_{out}(v,\textit{ColdMiss}) \,/\, \mathcal{W}_{total}(v)$ is sufficiently large (e.g., above $0.95$ in our experiments): first-touch traffic with negligible temporal locality. Its sector-count distribution is typically close to $S_{opt}(v)$, indicating spatially dense, well-coalesced accesses.

\noindent \underline{\textit{Small, Read-Only Broadcast Region.}} 
\label{pattern:broadcasting_constant_region}
A subgraph $G' \subseteq G$ is identified as exhibiting the \textit{Broadcasting Constant Region} pattern when it satisfies the following conditions in \metric{}: (1) every node $v \in V(G')$ exhibits the \textit{Broadcasting} pattern, and (2) the total cold-miss footprint of the subgraph fits within the capacity of the GPU read-only constant-cache path. Formally, $G'$ must satisfy the capacity constraint
\[
\textstyle\sum_{v \in V(G')} \mathcal{W}_{out}(v,\textit{ColdMiss}) \le \mathcal{C}_{const},
\]
where $\mathcal{C}_{const}$ is the effective capacity of the constant-cache hierarchy; the unique footprint of $G'$ is exactly its cold-miss traffic. Such a subgraph is a strong candidate for \texttt{\_\_constant\_\_} memory promotion, most beneficial when $RI(G')$ is high.

\section{THE \tool{} FRAMEWORK}\label{sec:overall_workflow}

\tool{} is a practical, general-purpose framework that reconstructs the \textit{Topological Distance} and \metric{} representations of \Cref{sec:methodology} from real-world GPU applications: it directly instruments fully optimized GPU binaries at runtime, collects fine-grained memory traces, and performs efficient offline analysis, requiring neither operating-system support nor source-level modifications.

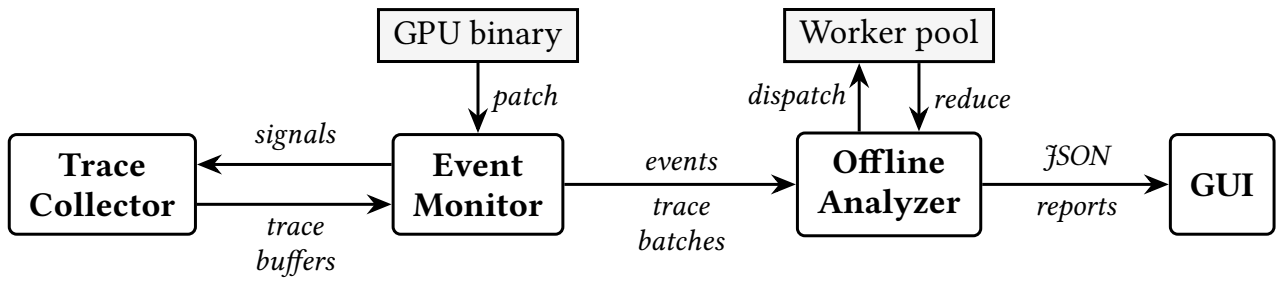
\begin{figure}[t]
    \centering
    \resizebox{\columnwidth}{!}{\begin{tikzpicture}[
  box/.style={draw, semithick, rounded corners=1.5pt, align=center,
              font=\scriptsize\bfseries, inner xsep=4pt, inner ysep=3pt,
              minimum height=20pt},
  aux/.style={draw, semithick, align=center, font=\scriptsize,
              inner xsep=3pt, inner ysep=2pt, fill=black!4},
  lbl/.style={font=\tiny\itshape, inner sep=1pt},
  arr/.style={-{Stealth[length=5pt,width=4pt]}, semithick},
  x=1cm, y=1cm,
]
\node[box] (tc)  at (0.0,0)  {Trace\\Collector};
\node[box] (em)  at (2.65,0) {Event\\Monitor};
\node[box] (oa)  at (5.55,0) {Offline\\Analyzer};
\node[box] (gui) at (7.9,0)  {GUI};
\node[aux] (bin) at (2.65,1.05) {GPU binary};
\node[aux] (wp)  at (5.55,1.05) {Worker pool};
\draw[arr] (bin) -- (em) node[lbl, midway, right=1.5pt] {patch};
\draw[arr] ([yshift=4pt]em.west) -- ([yshift=4pt]tc.east)
  node[lbl, midway, above=1.5pt] {signals};
\draw[arr] ([yshift=-4pt]tc.east) -- ([yshift=-4pt]em.west)
  node[lbl, midway, below=1.5pt, align=center] {trace\\buffers};
\draw[arr] (em.east) -- (oa.west)
  node[lbl, midway, above=1.5pt] {events}
  node[lbl, midway, below=1.5pt, align=center] {trace\\batches};
\draw[arr] ([xshift=-6pt]oa.north) -- ([xshift=-6pt]wp.south)
  node[lbl, midway, left=1.5pt] {dispatch};
\draw[arr] ([xshift=6pt]wp.south) -- ([xshift=6pt]oa.north)
  node[lbl, midway, right=1.5pt] {reduce};
\draw[arr] (oa.east) -- (gui.west)
  node[lbl, midway, above=1.5pt] {JSON}
  node[lbl, midway, below=1.5pt] {reports};
\end{tikzpicture}}
    \vspace{-0.62cm}
    \caption{The overview of \tool{} workflow.}
    \label{fig:workflow}
    \vspace{-0.42cm}
\end{figure}

\Cref{fig:workflow} illustrates the end-to-end workflow of \tool{}, which consists of four core components: \textit{Trace Collector}, \textit{Event Monitor}, \textit{Offline Analyzer}, and \textit{GUI}. 

\textbf{Trace Collector}\label{part:trace collector} patches fully optimized CUBINs at module load with SASS-level callbacks that collect runtime metrics and per-lane addresses on every memory instruction issue. We use the NVIDIA Compute Sanitizer API~\cite{SanitizerAPI} because it is a maintained CUDA-toolkit component that reports memory space (global, shared, or local) and access type at the callback boundary. Any instrumentation framework exposing the same facts (PC, address, access size, memory space, access type, block/warp/lane identity, and active mask), such as NVBit~\cite{villa2019nvbit}, could feed the same analysis pipeline.
The trace is organized at warp granularity, each buffer corresponding to one instruction instance of a single warp. Buffers are aggregated into batches and transferred to system memory for analysis through a producer--consumer queue between the GPU and CPU.

The \textit{Trace Collector} also supports asynchronous global-to-shared copies (\texttt{cp.async}, SASS \texttt{LDGSTS}), which modern kernels increasingly rely on for software pipelining. A single such instruction performs both a global read and a shared write, which would otherwise break the single-memory-space property of \metric{} nodes. \tool{} therefore splits each async-copy instance into two trace records attributed to synthetic PCs $pc$ and $pc+1$: since SASS instruction addresses are 16-byte aligned, the low bits are free to distinguish the global-read half from the shared-write half. Both records are tagged as async copies, so the global$\to$shared staging dataflow of pipelined kernels appears explicitly in \metric{}: subsequent shared-memory reads form reuse edges to the shared-write half.

\textbf{Event Monitor}\label{part:event monitor} handles driver events: module loading triggers instrumentation of whitelisted kernels; GPU memory (de)allocation triggers corresponding shadow-memory (de)allocation; and buffer-full notifications trigger trace transfer to system memory and analyzer invocation. It also tracks kernel launches and completion, exporting each completed kernel's profile as JSON keyed by a unique per-launch ID.

\textbf{Offline Analyzer}\label{part:offline analyzer} maintains GMEM and SMEM shadow-memory state on the host and reconstructs each kernel's \metric{} from the collected accesses.

Constructing an instruction-level, full-lifecycle dependency graph incurs substantial time and memory overhead: the last-access state must be consulted and updated on every dynamic memory access, over footprints that scale with the application's allocations. \tool{} keeps this high-fidelity analysis practical through three system-level designs.

\noindent\textit{(1) Direct-mapped, demand-paged shadow memory with an interleaved layout.} A hashmap keyed by address suffers from indirection, collisions, and rehashing that dominate at large footprints~\cite{zhao2025redsan}. Instead, \tool{} reserves a byte-wise host shadow region for every GPU allocation via \texttt{mmap}, making each lookup a single O(1) index computation. Demand paging commits physical pages only when the corresponding shadow bytes are first touched, and resetting a region is a page-table-level \texttt{madvise} rather than an O(size) clear. Tracking each access's metadata at its start address and every fourth byte thereafter suffices; interleaving the region into four byte-lane partitions (same-offset bytes packed contiguously) keeps these entries dense in host caches and commits physical pages in only one partition for aligned accesses.

\noindent\textit{(2) Compact lock-free shadow entries.} Each shadow entry packs the identity of the last access (flat block/warp/lane ID, PC offset, and a kernel-generation tag) into a single 64-bit word. A single atomic exchange then retrieves the previous access and installs the current one in one step, so concurrent workers never block on locks; the generation tag invalidates stale entries across kernel launches implicitly, so no per-launch reset of the shadow region is needed.

\noindent\textit{(3) Contention-free parallel workers.} Trace batches are partitioned by CTA ID across a pool of persistent worker threads. Each worker holds private analysis state, including graph-edge statistics, per-PC attributes, and CTA-local shared-memory shadow objects, so per-access statistics updates never contend, and shared-memory tracking requires no atomics at all because each CTA is processed by exactly one worker. Per-worker results are merged in a short post-batch sequential reduction, keeping all synchronization off the hot path.

Together, these designs keep the end-to-end overhead generally under 10$\times$ on production workloads (\Cref{sec:evaluation:overhead}).

\textbf{GUI} \label{part:gui} is a self-contained browser application that renders the JSON reports interactively at PC granularity; an optional \textit{Kernel Base PC} input converts PC offsets into absolute instruction addresses for direct cross-reference with NCU or nvdisasm output.

\noindent \textbf{Diagnostic Workflow.} \tool{} is designed to complement hardware profilers such as Nsight Compute (NCU). In the recommended workflow, developers first use NCU to identify bottleneck kernels and confirm memory-bound behavior through hardware counters (cache hit rates, stall breakdowns, throughput metrics). \tool{} is then applied to the suspicious kernel to diagnose the root cause: its \metric{} reveals which PCs are responsible, what data reuse patterns exist, and at which execution scope the reuse occurs. After applying an optimization, developers return to NCU to validate whether the hardware-level metrics have improved.

Our implementation is open-source, and its overhead characteristics have been independently reproduced by external users across multiple GPU platforms. It is available at \url{https://github.com/AccelProf/AccelProf}.

\section{EVALUATION}\label{sec:evaluation}
\begin{table*}[htbp]
  \centering
  \caption{Platform Configuration.}
  \label{tab:platform}
  \vspace{-0.45cm}
  \resizebox{0.85\textwidth}{!}{%
  \begin{tabular}{ccccccccc}
    \hline
    \multicolumn{1}{c}{\textbf{CPU}} & \multicolumn{1}{c}{\textbf{GPU}} & \multicolumn{1}{c}{\textbf{\#SM}} & \multicolumn{1}{c}{\textbf{GPU Memory}} & \multicolumn{1}{c}{\textbf{System}} & \multicolumn{1}{c}{\textbf{GPU Driver}} & \multicolumn{1}{c}{\textbf{CUDA Version}} & \multicolumn{1}{c}{\textbf{GCC}} & \multicolumn{1}{c}{\textbf{Compiler Options}} \\
    \hline
    AMD 5955WX (16c) & RTX 4090 & 128 & 24 GB GDDR6X & Linux 5.15.0 & 580.65.06 & 13.0 & 11.4.0 & -lineinfo -O3 -arch sm\_89 \\
    Intel(R) Xeon 8468 (48c) & H100 SXM & 132 & 80 GB HBM3 & Linux 5.15.0 & 550.163.01 & 12.4 & 11.4.0 & -lineinfo -O3 -arch sm\_90a \\
    \hline
  \end{tabular}%
  }%
\vspace{-0.45cm}
\end{table*}

We evaluate \tool{} on the two GPU workstations listed in \Cref{tab:platform}, with an \textit{RTX 4090 (sm\_89)} and an \textit{H100 (sm\_90a)}. The evaluation covers applications and benchmarks from multiple domains: \textit{Rodinia~\cite{che2009rodinia}}, \textit{XSBench~\cite{Tramm:wy}}, \textit{LULESH~\cite{LULESH2:changes}}, \textit{llama.cpp~\cite{llamacpp}} (commit \texttt{a8304b4d}), and \textit{PASTA~\cite{li2019pasta}}. Kernels with different template parameters or launch configurations are reported separately, since these choices can lead to substantially different performance behavior.
All reported inefficiencies were manually confirmed and optimized.

\subsection{Representation Fidelity and Stability}\label{sec:evaluation:fidelity}

\noindent\textbf{Pattern-recognition accuracy via microbenchmarks.} We first validate that \tool{} recognizes every memory behavior of \Cref{sec:recognized_patterns}, using 20 microbenchmark kernels that each implant one behavior whose \metric{} is analytically derivable, covering the six Topological Distance levels, the three redundancy variants, and known conflict and coalescing degrees. Every reported graph matches the derived expectation on schedule-invariant aggregates (edge labels and endpoint types, per-scope traffic volumes, sector and wavefront counts). Because the pattern rules are deterministic functions of these aggregates, every implanted behavior is classified correctly.

\noindent\textbf{Cross-validation against hardware counters.} By design, every executed warp-level memory instruction produces exactly one trace record with no sampling (two for asynchronous copies, one per memory space, matching how hardware counters attribute them; \Cref{sec:overall_workflow}), so the counts implied by \metric{} are directly comparable against independent hardware counters. We run every kernel of the paper's 19 applications (61 kernels, a strict superset of those optimized in \Cref{tab:speedup}) under both \tool{} and \ncu{} with identical inputs, excluding two spin-polling kernels whose load counts are not reproducible even across \ncu{} runs. On 58 of the remaining 59 kernels, every \tool{}-derived traffic count (global load/store requests, 32-byte sectors, shared-memory instructions, and atomic requests) agrees with \ncu{} to the last count, and the last kernel deviates by at most 0.03\%; shared-memory wavefront predictions agree within 2.5\% on all 34 shared-memory kernels, exactly on 19. Since replay-based counter collection has small error sources of its own, we read this agreement as mutual corroboration rather than either side serving as absolute ground truth. Counters corroborate only aggregate volumes; the reuse edges and scope labels \metric{} adds are covered by the microbenchmarks above, and since edge weights accumulate from the same per-access records whose totals match \ncu{}, the scope-decomposed traffic inherits this fidelity.

\noindent\textbf{Cross-platform stability.} \Cref{sec:pcdg} argues that \metric{} is determined by the program and its launch configuration, up to a single schedule-dependent tie-break at scope boundaries. We test this on both platforms of \Cref{tab:platform}: for each kernel and launch configuration run with identical inputs, we compare the byte-weighted distribution of traffic across the six reuse scopes by total variation distance (TVD). Across the 48 configurations with identical executed code and inputs, the distributions are identical on 19 and within 0.1\% TVD on 42; the mean is 0.08\% and the maximum 1.75\%. Every deviation above 0.1\% is a compensating shift between adjacent scopes (PASTA's \textit{mttkrp} relabels 1.75\% of its bytes from intra-grid to intra-block), exactly the predicted tie-break: warp scheduling can change which same-data candidate is the temporally nearest reuse, and thus the boundary label, but not the volume of reuse. The remaining differences track genuine platform differences: llama.cpp's \textit{parallelize\_cols} launches one block per SM, so its grid grows from 128 to 132 blocks, yet its signature matches within 0.1\%; a CUB-based \textit{cumsum} kernel selects different code on sm\_89 and sm\_90a and yields a correspondingly different signature (6\% TVD), which \tool{} reports faithfully.

\subsection{Actionability of \metric{}}\label{sec:evaluation:patterns_speedup}
\metric{} provides a logical view of memory behavior, from which some observed patterns can be identified as inefficient. However, not all such inefficiencies are equally actionable: some may be largely absorbed by hardware mechanisms or compiler optimizations, while others remain performance-critical in the final execution. Across diverse benchmarks and real applications, we use \tool{} to locate these actionable inefficiency patterns and validate targeted fixes with only minor code modifications, typically fewer than 10 lines.

Because the graph diagnoses \emph{structural} mismatch (reuse served by a mechanism that does not match its scope), the primary validation criterion is structural as well: whether the guided fix removes exactly the waste the graph attributes, on the meter of the mechanism involved. We therefore measure each fix by its reduction in memory-instruction count for register-serviceable reuse, in shared-memory wavefronts for bank conflicts, and in global-memory (L2--DRAM) traffic for grid-scope contention; these quantities are deterministic, independent of warp scheduling, and verifiable against hardware counters at the exactness established in \Cref{sec:evaluation:fidelity}. Kernel speedup, measured as the ratio of kernel cycles (\Cref{tab:speedup}), is reported alongside as corroboration, but it is an incomplete indicator of a memory fix in isolation: the removed operations may lie off the critical path or be covered by latency hiding and concurrent instruction streams, so a structurally confirmed fix can leave runtime nearly unchanged on one platform while its eliminated traffic still matters for contention, energy, and other occupancy regimes.

As summarized in \Cref{tab:speedup}, the guided fixes eliminate up to $90.18\%$ of memory instructions across the table's six optimization categories ($28.7\%$ on average over the memory-instruction entries of both platforms, excluding the wavefront- and traffic-validated rows), with kernel speedups reaching 2.84$\times$. The two exceptions follow their own meters: the bank-conflict fixes restructure shared-memory addressing rather than remove instructions, and their starred cells report wavefront reductions of 41--70\%; XSBench's dagger-marked fix cuts global-memory traffic by 29.13\% while leaving instruction counts essentially unchanged.

\newcommand{\thinhline}{\noalign{\hrule height 0.1pt}}
\newcommand{\thickhline}{\noalign{\hrule height 1.2pt}}

\begin{table*}[t]
\centering
\caption{Kernel and memory performance comparison on RTX 4090 and H100. Speedup is original over optimized kernel cycles (frequency-independent), averaged over 10 Nsight Compute runs; memory instruction reduce counts load, store, and atomic instructions across local, shared, and global memory. XSBench's grid-scope contention exceeds the H100 L2 (50\,MB); its fix is instead evaluated on an RTX 5090 (96\,MB L2), with speedup and traffic reduction similar to the RTX 4090 results reported here. $^{*}$Bank-conflict fixes do not remove memory instructions (counts change by at most 0.39\%); starred cells report the reduction in serialized shared-memory wavefronts instead. $^{\dagger}$Fixes targeting grid-scope contention are validated by global-memory (L2--DRAM) traffic; XSBench's drops by 29.13\%.
}
\label{tab:speedup}
\vspace{-0.4cm}
\setlength{\tabcolsep}{4pt}
\renewcommand{\arraystretch}{1.15}
\resizebox{0.98\textwidth}{!}{%
\begin{tabular}{ccccccc}
\thickhline
\multirow{3}{*}{\textbf{Application}} & \multirow{3}{*}{\textbf{Kernel Name}} & \multirow{3}{*}{\textbf{Inefficiency Pattern}} & \multicolumn{2}{c}{\textbf{RTX 4090}} & \multicolumn{2}{c}{\textbf{H100}} \\
\cline{4-5}\cline{6-7}
& & & \textbf{Kernel} & \textbf{Mem Inst.} & \textbf{Kernel} & \textbf{Mem Inst.} \\
& & & \textbf{Speedup} & \textbf{Reduce} & \textbf{Speedup} & \textbf{Reduce} \\
\thickhline
\multirow{2}{*}{\textbf{rodinia/backprop}} & bpnn\_layerforward\_CUDA & redundant & 1.02$\times$ & 12.03\% & 0.98$\times$ & 12.03\% \\
 & bpnn\_adjust\_weights\_cuda & redundant & 1.55$\times$ & 33.33\% & 1.03$\times$ & 33.33\% \\
\thinhline
\textbf{rodinia/bfs} & Kernel & L2 contention / redundant & 1.08$\times$ & 7.56\% & 1.07$\times$ & 7.03\% \\
\thinhline
\textbf{rodinia/hotspot} & calculate\_temp & abuse SMEM & 1.04$\times$ & 31.65\% & 1.05$\times$ & 31.65\% \\
\thinhline
\multirow{2}{*}{\textbf{rodinia/huffman}} & vlc\_encode\_kernel\_sm64huff & bank conflict & 1.04$\times$ & 50.37\%$^{*}$ & 1.01$\times$ & 50.32\%$^{*}$ \\
 & prescan$<$(bool)1, (bool)0$>$ & bank conflict & 1.01$\times$ & 42.60\%$^{*}$ & 1.02$\times$ & 41.80\%$^{*}$ \\
\thinhline
\multirow{2}{*}{\textbf{rodinia/lud}} & lud\_diagonal & bank conflict & 1.04$\times$ & 41.35\%$^{*}$ & 1.03$\times$ & 41.35\%$^{*}$ \\
 & lud\_perimeter & bank conflict & 1.15$\times$ & 69.83\%$^{*}$ & 1.14$\times$ & 69.83\%$^{*}$ \\
\thinhline
\multirow{2}{*}{\textbf{rodinia/nw}} & needle\_cuda\_shared\_1 & bank conflict & 1.07$\times$ & 68.85\%$^{*}$ & 1.04$\times$ & 68.85\%$^{*}$ \\
 & needle\_cuda\_shared\_2 & bank conflict & 1.07$\times$ & 68.85\%$^{*}$ & 1.04$\times$ & 68.85\%$^{*}$ \\
\thinhline
\multirow{3}{*}{\textbf{rodinia/srad\_v1}} & reduce & redundant & 2.62$\times$ & 90.18\% & 2.08$\times$ & 90.18\% \\
& srad & constant & 1.01$\times$ & 14.28\% & 1.00$\times$ & 14.29\% \\
& srad2 & constant & 0.97$\times$ & 9.09\% & 1.00$\times$ & 9.09\% \\
\thickhline
\textbf{llama.cpp/cumsum} & cumsum\_kernel & redundant & 0.99$\times$ & 21.14\% & 1.05$\times$ & 21.14\% \\
\thinhline
\multirow{3}{*}{\textbf{llama.cpp/rms\_norm}} & rms\_norm\_f32$<$(int)1024, (bool)0, (bool)0$>$$\lll$(5,4,3), (1024,1,1)$\ggg$ & redundant & 1.08$\times$ & 20.25\% & 1.06$\times$ & 20.25\% \\
 & rms\_norm\_f32$<$(int)256, (bool)0, (bool)0$>$$\lll$(2,2,1),(256,1,1)$\ggg$ & redundant & 1.01$\times$ & 15.13\% & 1.02$\times$ & 15.13\% \\
 & rms\_norm\_f32$<$(int)256, (bool)0, (bool)0$>$$\lll$(5,4,3),(256,1,1)$\ggg$ & redundant & 1.01$\times$ & 9.09\% & 1.00$\times$ & 9.09\% \\
\thinhline
\multirow{2}{*}{\textbf{llama.cpp/rms\_norm\_back}} & rms\_norm\_back\_f32$<$(int)32$>$ & redundant & 1.00$\times$ & 40.00\% & 0.99$\times$ & 40.00\% \\
 & rms\_norm\_back\_f32$<$(int)1024$>$ & redundant & 1.07$\times$ & 22.53\% & 1.09$\times$ & 22.53\% \\
\thinhline
\multirow{6}{*}{\textbf{llama.cpp/soft\_max}} & soft\_max\_f32$<$(bool)1, (int)1024, (int)1024, \_\_half$>$$\lll$(16,1,1),(1024,1,1)$\ggg$ & redundant / abuse SMEM & 1.03$\times$ & 33.33\% & 1.00$\times$ & 33.33\% \\
 & soft\_max\_f32$<$(bool)1, (int)1024, (int)1024, \_\_half$>$$\lll$(1024,1,1),(1024,1,1)$\ggg$ & redundant / abuse SMEM & 1.06$\times$ & 33.33\% & 1.02$\times$ & 33.33\% \\
 & soft\_max\_f32$<$(bool)1, (int)1024, (int)1024, float$>$$\lll$(16,1,1),(1024,1,1)$\ggg$ & redundant / abuse SMEM & 1.05$\times$ & 34.29\% & 1.00$\times$ & 34.29\% \\
 & soft\_max\_f32$<$(bool)1, (int)1024, (int)1024, float$>$$\lll$(1024,1,1),(1024,1,1)$\ggg$ & redundant / abuse SMEM & 1.05$\times$ & 34.29\% & 1.02$\times$ & 34.29\% \\
 & soft\_max\_f32$<$(bool)1, (int)32, (int)32, float$>$$\lll$(2,32,1),(32,1,1)$\ggg$ & abuse SMEM & 1.03$\times$ & 16.67\% & 1.01$\times$ & 16.67\% \\
 & soft\_max\_f32$<$(bool)1, (int)32, (int)32, \_\_half$>$$\lll$(2,32,1),(32,1,1)$\ggg$ & abuse SMEM & 0.97$\times$ & 20.83\% & 0.97$\times$ & 20.00\% \\
\thickhline
\multirow{3}{*}{\textbf{LULESH}} & ApplyMaterialPropertiesAndUpdateVolume\_kernel & constant & 1.00$\times$ & 29.58\% & 1.01$\times$ & 29.58\% \\
 & CalcTimeConstraintsForElems\_kernel & redundant & 1.01$\times$ & 26.36\% & 1.13$\times$ & 51.71\% \\
 & CalcMinDtOneBlock & redundant & 1.22$\times$ & 62.07\% & 1.10$\times$ & 62.07\% \\
\thickhline
\textbf{XSBench} & xs\_lookup\_kernel\_baseline & L2 contention  & 1.03$\times$ & 0\%$^{\dagger}$ & -- & -- \\
\thickhline
\multirow{2}{*}{\textbf{PASTA}} & spt\_MTTKRPKernelRankSplitNnz4DOneKernel & L1 contention / constant & 1.00$\times$ & 27.53\% & 1.00$\times$ & 27.53\% \\
 & spt\_TTMRankRBNnzKernelSM &  L1 contention / abuse SMEM & 2.84$\times$ & 32.04\% & 1.67$\times$ & 32.04\%\\
\thickhline

\end{tabular}%
\vspace{-0.4cm}
}
\end{table*}

\subsection{Overhead}\label{sec:evaluation:overhead}
We also report the end-to-end overhead of \tool{} on the RTX 4090 platform. No sampling is used: every instrumented memory instruction is traced and analyzed, so the reported overhead reflects the full cost of exact dependency reconstruction. As shown in \Cref{tab:overhead}, \tool{} remains practical for high-fidelity analysis on real workloads, with overhead generally under $10\times$ (median 3.2$\times$). For calibration, \redsan{}, which tracks the same last-access state through a serial hashmap, reports 34.36$\times$ on \textit{rodinia/backprop} where \tool{} incurs 4.80$\times$ (\Cref{sec:related_work}). The two major outliers are \textit{rodinia/srad\_v1} (53.75$\times$), which issues 502 launches of six small kernels so instrumentation dominates, and XSBench (25.28$\times$), whose memory-instruction volume implies a pre-cache analysis footprint of about 415\,GB. A kernel whitelist lets users bound analysis effort on such complex applications.

\begin{figure}[!ht]
    \centering
    \vspace{-0.2cm}
    \includegraphics[width=0.9\linewidth]{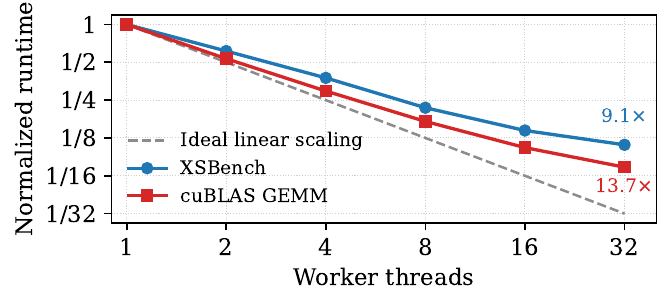}
    \vspace{-0.3cm}
    \caption{Analyzer runtime with different worker-thread counts, normalized to one worker, for XSBench (GMEM-only) and cuBLAS GEMM (SMEM-heavy).}\vspace{-0.2cm}
    \Description{Scalability plot comparing normalized analysis runtime across different worker-thread counts for two workload types.}
    \label{fig:scalability}
    \vspace{-0.3cm}
\end{figure}

We further evaluate the scalability of the parallel offline analyzer under different worker-thread counts using two representative workloads: XSBench (input \textit{-s large -m event -l 17000000}), whose traffic is global-memory only, and a cuBLAS $4096\times4096$ GEMM, where $82\%$ of the accesses target shared memory. As shown in \Cref{fig:scalability}, the analyzer runs 9.1$\times$ and 13.7$\times$ faster than the single-worker configuration at 32 threads, respectively. Notably, the host CPU has only 16 physical cores (\Cref{tab:platform}), so the 16$\to$32-thread step runs entirely on SMT yet still gains another 30\% and 43\%; the 13.7$\times$ speedup corresponds to 86\% of the physical core count.

\begin{table}[!ht]
\centering
\caption{Overhead comparison between the original and \tool{}-enabled executions on the RTX 4090 platform in \Cref{tab:platform} with 32 worker threads.}
\vspace{-0.4cm}
\label{tab:overhead}
\small
\renewcommand{\arraystretch}{1.15}
\resizebox{0.95\linewidth}{!}{%
\begin{tabular}{@{}cccc@{}}
\toprule
\textbf{Application Name} & \textbf{origin (ms)} & \textbf{\tool{}\_enabled (ms)} & \textbf{Overhead} \\
\midrule
\textbf{rodinia/backprop} & 204 & 979 & 4.80$\times$ \\
\textbf{rodinia/bfs} & 2912 & 9408 & 3.23$\times$ \\
\textbf{rodinia/hotspot} & 314 & 1013 & 3.23$\times$ \\
\textbf{rodinia/srad\_v1} & 206 & 11073 & 53.75$\times$ \\
\textbf{llama.cpp/cumsum} & 460 & 1445 & 3.14$\times$ \\
\textbf{llama.cpp/rms\_norm} & 511 & 1417 & 2.77$\times$ \\
\textbf{llama.cpp/rms\_norm\_back} & 433 & 1212 & 2.80$\times$ \\
\textbf{llama.cpp/soft\_max} & 4824 & 10501 & 2.18$\times$ \\
\textbf{XSBench} & 4432 & 112042 & 25.28$\times$ \\
\textbf{LULESH} & 395 & 3809 & 9.64$\times$ \\
\textbf{pasta/mttkrp} & 2080 & 13592 & 6.53$\times$ \\
\textbf{pasta/ttm\_hicoo} & 3492 & 5094 & 1.46$\times$ \\
\bottomrule
\end{tabular}%
}
\vspace{-0.35cm}
\end{table}

\section{CASE STUDIES AND DISCUSSION}\label{sec:case_studies}
This section demonstrates \metric{} as an analysis substrate rather than a fixed report: a recurring redundancy diagnosed and fixed in real applications (\Cref{sec:case_studies:ai_redundancy}), a path from these developer-driven analyses to profile-guided optimization (\Cref{sec:pgo_direction}), and a discussion of what \metric{} carries beyond performance, and its limitations (\Cref{sec:case_studies:discussion}).

\subsection{A Common Redundancy Across AI Libraries}\label{sec:case_studies:ai_redundancy}
During evaluation of \tool{}, we identified a recurring redundancy pattern in multi-pass operators (\textit{softmax}, \textit{layer\_norm}, and \textit{rms\_norm}, etc.) across multiple AI libraries (including llama.cpp~\cite{llamacpp}, ONNXruntime~\cite{onnxruntime}, PyTorch~\cite{pytorch}, llm.c~\cite{karpathy2024llmc}, and Triton~\cite{triton} tutorial). These multi-pass operators typically require at least two scans of the same input: the first pass computes a global statistic (e.g., maximum) and broadcasts it to threads in the same block, while the second pass re-scans the input for final outputs using that statistic.

\begin{figure}
\begin{lstlisting}[caption={A softmax kernel listing showing max-reduction, exponentiation, and normalization.}, label={case_code:softmax}]
template <...int ncols_template...>
static __global__ void soft_max_f32(...) {
  const int ncols = ncols_template == 0 ? p.ncols : ncols_template;
  float * @vals@ = use_shared ? buf_iw + WARP_SIZE : dst;
  ...
  for (int col0 = 0; col0 < ncols; col0 += block_size) {
    const int col = col0 + tid; // col is thread unique
    const float val = x[col]*p.scale + ...;
    @vals[col]@ = val; max_val = max(max_val, val);
  }
  max_val = block_reduce<MAX,...>(max_val, buf_iw);
  for (...) { // same thread-unique col indexing
    const float val = expf(@vals[col]@ - max_val);
    tmp += val; @vals[col]@ = val;
  }
  tmp = block_reduce<SUM,...>(tmp, buf_iw);
  for (...) dst[col] = @vals[col]@ * inv_sum;
}
\end{lstlisting}
\vspace{-0.5cm}
\end{figure}

In \Cref{case_code:softmax}, when the template parameter \textit{use\_shared} is true, the \textit{vals} pointer is placed in shared memory but accessed in a thread-private order. \Cref{fig:dependency_graph_example} in \Cref{sec:pcdg} is the \metric{} of this kernel, and the misuse is legible as graph structure: the four \texttt{SHARED} PCs accessing \textit{vals} form the leftmost closed subgraph (Group~1), and every edge inside it is \textit{intra\_thread}: each element staged in shared memory is written and read back only by the thread that produced it. The same figure supplies the control case: the second closed subgraph (Group~2), formed by the broadcast buffer \textit{buf\_iw}, carries the \textit{intra\_block} and \textit{intra\_warp} edges that justify block-scope storage. Both subgraphs live in the same shared-memory allocation (\textit{vals} is \textit{buf\_iw}\,+\,\texttt{WARP\_SIZE}), yet only one exhibits cross-thread reuse; this distinction is invisible to bandwidth- or transaction-level metrics and is precisely the scope information \metric{} attaches to every reuse pair. Thread-private data held in block-scope storage forfeits the core purpose of shared memory (low-latency, software-managed sharing within a thread block), introduces redundant memory instructions, and reduces the available L1 capacity, since shared memory is carved out of the same physical array.

Another important limitation leading to redundancy comes from the statically unknown loop bound (\textit{ncols} in this example). Because the compiler cannot determine how many elements each thread will process, it may conservatively generate redundant loads or spill values to local memory, even when register pressure is not the limiting factor.

This scope misplacement is not specific to AI libraries. In the Rodinia suite, \textit{hotspot} allocates the shared-memory objects \texttt{power\_\allowbreak on\_device} and \texttt{temp\_t} that are each touched by only one thread, and \textit{backprop} stages weights in a shared-memory array that each thread writes and then immediately reads back itself.

\subsection{Toward Profile-Guided Memory Optimization}
\label{sec:pgo_direction}

While \tool{} currently informs manual restructuring, nothing about consuming \metric{} is inherently manual. The patterns in \Cref{sec:recognized_patterns} are mechanical predicates over the graph, and several of the transformations they suggest (register promotion for redundant accesses, \texttt{\_\_constant\_\_} promotion for broadcast regions, shared-memory demotion for thread-private SMEM) are mechanical. A natural next step is to close the loop: treat \metric{} as a memory-placement profile fed back into compilation, in the same way edge and path profiles guide code layout and inlining in classical profile-guided optimization. Because the representation is stable across GPU targets, such profiles are reusable across hardware generations; and because it is extracted from optimized binaries, the feedback reflects the code the compiler emitted rather than a source-level approximation.

The same information could also flow downward to hardware. Semantic-aware access annotations (e.g., thread-unique, warp-broadcast, streaming) derived from \metric{} could be lowered by the compiler onto specialized hardware paths (constant caches, L2 eviction policies, or automated shared-memory mapping), replacing the conservative assumptions that today lead to cache thrashing and wasted bandwidth.

\subsection{Discussion}\label{sec:case_studies:discussion}

\noindent\textbf{Beyond performance.}\label{sec:case_studies:beyond_inefficiency}
Although \metric{} is designed for a performance tool, its edges carry correctness information. A cross-thread reuse edge with a write on either endpoint is a pair of conflicting accesses, and its Topological Distance names the synchronization scope that must order the pair: any such edge not separated by a barrier of at least that scope is a potential race. Race candidates thus become an offline query over the same profile. The graph in \Cref{fig:dependency_graph_example} contains one: the cross-thread write-after-read edges from \texttt{0xa30} to \texttt{0x880} in Group~2 record a block-level reduction overwriting the broadcast buffer a previous reduction is still reading, with no intervening \texttt{\_\_syncthreads()}; the race is genuine and was later fixed upstream (commit \texttt{9bd4c09e}). Because the store's shadow comparison meets the earlier read under every interleaving, the pair is captured on every run rather than only under a racy schedule, already attributed to both PCs and their scope. The same rule read in reverse degenerates to an inefficiency check: a barrier spanned by no cross-thread reuse edge orders nothing, which is how \metric{} exposes the removable \texttt{\_\_syncthreads()} in \textit{rodinia/backprop} (\Cref{tab:speedup}). We deliberately stop short of building a race checker; dedicated GPU race detectors address the general problem~\cite{eizenberg2017barracuda, kamath2021iguard}. The observation is narrower: the conflicting-pair structure that constitutes the expensive half of such analyses is already present in the profile \tool{} extracts once, so extending the analysis surface means writing an offline query, not new instrumentation.

\noindent\textbf{Limitations.}\label{sec:limitations}
As a dynamic analysis, \tool{} inherits the classical boundaries of profiling. First, the reconstructed \metric{} reflects the executed control flow and data distribution: reuse behavior that a given input does not exercise remains invisible, and strongly input-dependent kernels should be profiled under representative inputs. Second, the analysis is diagnostic rather than prescriptive: the patterns of \Cref{sec:recognized_patterns} name the responsible instructions, scopes, and applicable hardware mechanisms, but applying a fix still involves developer judgment (e.g., weighing register pressure or occupancy), and closing this loop automatically is the profile-guided direction outlined above. Third, while the representation itself is vendor-neutral, our implementation targets NVIDIA GPUs through the Compute Sanitizer API; porting requires a substrate that reports the same per-lane facts (\Cref{sec:overall_workflow}). Finally, tracing every access has a cost that grows with dynamic memory-instruction count: most workloads stay under 10$\times$, but launch-heavy or extremely instruction-heavy applications reach 25--54$\times$ and rely on the kernel whitelist to bound analysis effort (\Cref{sec:evaluation:overhead}).

\section{CONCLUSION}

We presented \textit{Topological Distance} and \metric{}, a runtime representation of GPU memory reuse structure defined over execution scopes, together with \tool{}, which reconstructs it from optimized binaries at profiler-class overhead. \metric{} reports how data is actually shared across instructions and execution scopes, the facts that optimization decisions consume. Across HPC and AI workloads on two GPU generations, the representation matched hardware counters, remained stable across platforms, and guided fixes that eliminated up to 90.18\% of memory instructions. We see this as a step toward reuse structure as a first-class program profile: today's manual-fix analyses are already mechanical rules over the graph, pointing toward profile-guided memory optimization.

\bibliographystyle{ACM-Reference-Format}
\bibliography{reference.bib}

\end{document}